\documentclass[prd,twocolumn,a4paper,superscriptaddress,floatfix]{revtex4}
\usepackage{graphicx}
\usepackage{bbm}
\usepackage[all]{xy}
\usepackage{amsmath}
\usepackage{amssymb}
\usepackage{epstopdf}

\def\be{\begin{equation}}
\def\ee{\end{equation}}
\newcommand{\bq}{\begin{eqnarray}}
\newcommand{\eq}{\end{eqnarray}}
\newcommand{\bes}{\begin{subequations}}
\newcommand{\ees}{\end{subequations}}

\def\ben{\begin{eqnarray}}
\def\een{\end{eqnarray}}
\def\ba{\begin{array}}
\def\ea{\end{array}}

\begin{document}
\newcommand{\half}{{\textstyle\frac{1}{2}}}
\allowdisplaybreaks[3]
\def\a{\alpha}
\def\b{\beta}
\def\g{\gamma}\def\G{\Gamma}
\def\d{\delta}\def\D{\Delta}
\def\ep{\epsilon}
\def\et{\eta}
\def\z{\zeta}
\def\t{\theta}\def\T{\Theta}
\def\l{\lambda}\def\L{\Lambda}
\def\m{\mu}
\def\f{\phi}\def\F{\Phi}
\def\n{\nu}
\def\p{\psi}\def\P{\Psi}
\def\r{\rho}
\def\s{\sigma}\def\S{\Sigma}
\def\ta{\tau}
\def\x{\chi}
\def\o{\omega}\def\O{\Omega}
\def\k{\kappa}
\def\pa {\partial}
\def\ov{\over}
\def\br{\\}
\def\ud{\underline}

\newcommand{\sech}{{\rm sech}}
\newcommand{\LL}{{\cal L}}
\newcommand{\XX}{{\cal X}}
\newcommand{\LX}{{\cal L}_X}
\newcommand{\LXone}{{\cal L}_{X_1}}
\newcommand{\LXtwo}{{\cal L}_{X_2}}
\newcommand{\LXN}{{\cal L}_{X_N}}
\newcommand{\LXS}{{\cal L}_{X_S}}
\newcommand{\LXij}{{\cal L}_{X_{ij}}}
\newcommand{\LXP}{{\cal L}_{X_{\phi}}}
\newcommand{\LXQ}{{\cal L}_{X_{\chi}}}
\newcommand{\LP}{{\cal L}_{\phi}}
\newcommand{\LQ}{{\cal L}_{\chi}}
\newcommand{\LPi}{{\cal L}_{\phi_i}}
\newcommand{\LXX}{{\cal L}_{XX}}
\newcommand{\LXXS}{{\cal L}_{X_S X_S}}
\newcommand{\LXSP}{{\cal L}_{X_S\phi}}

\newcommand\lsim{\mathrel{\rlap{\lower4pt\hbox{\hskip1pt$\sim$}}
    \raise1pt\hbox{$<$}}}
\newcommand\gsim{\mathrel{\rlap{\lower4pt\hbox{\hskip1pt$\sim$}}
    \raise1pt\hbox{$>$}}}
\newcommand\esim{\mathrel{\rlap{\raise2pt\hbox{\hskip0pt$\sim$}}
    \lower1pt\hbox{$-$}}}

\title{Localized D-dimensional global k-defects}

\author{P.P. Avelino}
\affiliation{Centro de F\'{\i}sica do Porto, Rua do Campo Alegre 687, 4169-007 Porto, Portugal}
\affiliation{Departamento de F\'{\i}sica da Faculdade de Ci\^encias
da Universidade do Porto, Rua do Campo Alegre 687, 4169-007 Porto, Portugal}
\affiliation{Departamento de F\'{\i}sica, Universidade Federal da Para\'{\i}ba
58051-970 Jo\~ao Pessoa, Para\'{\i}ba, Brasil}
\author{D. Bazeia}
\affiliation{Departamento de F\'{\i}sica, Universidade Federal da Para\'{\i}ba
58051-970 Jo\~ao Pessoa, Para\'{\i}ba, Brasil}
\author{R. Menezes}
\affiliation{Centro de F\'{\i}sica do Porto, Rua do Campo Alegre 687, 4169-007 Porto, Portugal}
\affiliation{Departamento de F\'{\i}sica da Faculdade de Ci\^encias
da Universidade do Porto, Rua do Campo Alegre 687, 4169-007 Porto, Portugal}
\affiliation{Departamento de Ci\^encias Exatas, Universidade Federal da Para\'{\i}ba, 58297-000 Rio Tinto PB, Brazil}
\affiliation{Departamento de F\'{\i}sica, Universidade Federal de Campina Grande, 58109-970, Campina Grande, Para\'{\i}ba, Brazil}
\author{J.G.G.S. Ramos}
\affiliation{Departamento de F\'{\i}sica, Universidade Federal da Para\'{\i}ba
58051-970 Jo\~ao Pessoa, Para\'{\i}ba, Brasil}

\begin{abstract}

We explicitly demonstrate the existence of static global defect solutions of arbitrary dimensionality whose energy does not diverge at spatial infinity, by considering maximally symmetric solutions described by an action with non-standard kinetic terms in a $D+1$ dimensional  Minkowski space-time. We analytically determine the defect profile both at small and large distances from the defect centre. We study the stability of such solutions and discuss possible implications of our findings, in particular for dark matter and charge fractionalization in graphene.

\end{abstract} 
\maketitle

\section{Introduction}

Scalar fields are expected to play a major cosmological role both at recent and early times. The most popular explanation for the current acceleration of the Universe (see, for example, \cite{Frieman:2008sn,Komatsu:2010fb}) relies on a nearly homogeneous dark energy component violating the strong energy condition, described by a minimally coupled scalar field, known as quintessence \cite{Zlatev:1998tr}. Scalar fields are also a crucial ingredient in symmetry breaking phase transitions which may have led to the generation of topological defect networks in the early Universe \cite{1994csot.book.....V}. Although domain wall networks seem to be definitely ruled out as a relevant dark energy component \cite{PinaAvelino:2006ia,Avelino:2008ve,Sousa:2009is}, other topological defects, such as cosmic strings, may still provide an important but subdominant (at least on large scales) contribution as seeds of large scale structure \cite{Wu:1998mr,Avelino:2003nn,Bevis:2007gh}. On the other hand, cosmological scalar fields are also expected to play an essencial role in the context of primordial inflation, providing a solution to some of the most fundamental cosmological enigmas (see, for example, \cite{Olive:1989nu} and references therein).

Although most studies consider minimally coupled scalar fields described by standard Lagrangians, in recent years there has been an increasing number of papers dealing with non-canonical kinetic terms. Such scalar fields are also known as k-fields and they may be relevant both in the context of inflation \cite{1999PhLB..458..209A} and dark energy \cite{2000PhRvL..85.4438A,Gorini:2003wa}, in particular in the context of unified scenarios \cite{Avelino:2008cu,Bose:2008ew}, and may be associated to space-time variations of fundamental couplings \cite{Avelino:2008dc}. They may also lead to topological defects whose properties are rather different from the standard ones \cite{Babichev:2006cy,Bazeia:2007df,Jin:2007fz,Adam:2007ij,Sarangi:2007mj,Babichev:2007tn,Adam:2008ck,Bazeia:2008tj,BlancoPillado:2008cp,Babichev:2008qv,Adam:2008rf,Liu:2009eh,Adam:2009px,Bazeia:2009db,Bazeia:2010vb,Bazeia:2010wr}. 

In this paper we consider static global k-defect solutions described by a simple generalization of the standard scalar field Lagrangian.
We investigate the stability of these solutions against radial stretching and demonstrate the existence of static global defect solutions whose energy does not diverge at spatial infinity. We show that this may lead to important changes to the evolution of defect networks and we discuss the possible role of global k-monopoles as a dark matter source. In two spatial dimensions the present investigation may also be relevant to the study of the fermionic charge fractionalization in graphene. Although the classical static vortex solution is central to the idea of fractionalization \cite{chamon}, the assumption of standard dynamics makes the energy of the vortex ill-defined in the case of a global symmetry. To cure this problem an alternative model was proposed  \cite{jackiw}, with the inclusion of a gauge potential which makes the symmetry local, thus leading to vortex solutions with finite energy. As we will show below, the modification of the dynamics which we suggest in this work leads to a finite vortex energy even when the symmetry is global without the need for the gauge vector potential.

Throughout the work, we will assume the metric signature $[-,+,...,+]$ and the calculations will be done using units in which $c=\hbar=1$. The Einstein summation convention will be used when a latin or greek index variable appears twice in a single term, once in an upper (superscript) and once in a lower (subscript) position. Except if stated otherwise, greek and latin indices take the values $0,...,D$ and $1,...,D$, respectively.

\section{Standard global defects}

We start by considering a real scalar field multiplet $\{\phi^1, ..., \phi^D\}$ in a $D+1$ dimensional Minkowski space-time described by the action $S=\int \LL \, d^{D+1}x$ with
\be
\label{standardL}
\LL =X - V(\phi^a)\,.
\ee
Here $X=-\delta_{ab} \phi^a_{,\mu} \phi^{b,\mu}/2$, $\delta_{ab}$ being the Kronecker delta ($\delta_{ab}=1$ if $a=b$ and $\delta_{ab}=0$ if $a \neq b$), a comma denotes a partial derivative and $V \ge 0$. The energy-momentum tensor for this model is given by
\be
T_{\mu\nu}=\delta_{ab} \phi^a_{,\mu} \phi^b_{,\nu} + g_{\mu\nu}\LL\,,
\ee
and the total energy can be computed as $E=\int d^D x\, T_{00}$. The possible existence of stable solutions with finite energy for $D > 1$ was discarded by Derrick and Hobart \cite{0370-1328-82-2-306,1964JMP.....5.1252D} assuming a standard Lagrangian given by Eq. (\ref{standardL}). In this case the gradient and potential contributions to the total energy, $E=K+U$, are given by
\be\label{EGandEV}
K=-\int d^Dx  \, X\,, \, \,\,\,\, U=\int d^Dx \, V(\phi^a)\,,
\ee
and a simple scaling argument was used to demonstrate that any static solution of this kind with finite $E$ would tend to collapse if $D>1$. Still, the existence of static global string and monopole solutions is not forbidden since these are cases for which the gradient energy, $K$, and consequently the total energy, $E$,  formally diverges. Of course, in physically realistic situations there will always be a cutoff at some energy scale \cite{1992NuPhB.375..665P}. For example, in a cosmological context, the mild logarithmic divergence in the energy of a global string has a cutoff due to the finite characteristic length of the string network.

In this paper we consider maximally symmetric static solutions in $D+1$ Minkowski space-times 
given by
\be
\label{ans}
\phi^a=\frac{x^a}{r} H(r)\,,
\ee 
with $r^2=x_a x^a$. In this case $X$ is a function of $r$ alone,
\be
X=-\frac12 \left((H_{,r})^2+(D-1)\frac{H^2}{r^2}\right)\,.
\ee
At large distances from the defect core $H \to 1$ so that
\be
X \to -\frac{D-1}{2r^2}\,,
\ee
and consequently $K$ and $E$ diverge in the $r \to \infty$ limit for $D > 1$. The analysis can be generalized to describe static p-brane solutions in a $N+1$ dimensional Minkowski space-time with $N > D$ by assuming that the scalar field multiplet is independent of the additional $p=N-D$ space-time coordinates. As a consequence such p-branes are featureless along the extra $p$ dimensions.

\section{Localized global k-defects}

Consider a real scalar field multiplet $\{\phi^1, ..., \phi^D\}$ in a D+1 dimensional Minkowski space-time described by a generic  Lagrangian 
\be
\label{gen}
\LL = \LL(\phi^a,X^{bc})\,,
\ee
where $X^{bc}=-\phi^b_{,\mu} \phi^{c,\mu}/2$. The energy-momentum tensor for this model is given by
\be
T_{\mu\nu}=\LL_{,X^{ab}}  \phi^a_{,\mu} \phi^b_{,\nu} + g_{\mu\nu}\LL\,,
\ee
and the equation  of motion is given by
\be
\label{eqmotion}
\frac{1}{\sqrt{-g}}\, \left(\sqrt{-g}\mathcal{L}_{,X}\phi^{,\mu}\right)_{,\mu}=-\mathcal{L}_{,\phi}\,.
\ee
In this paper we shall consider static solutions with
\be
E=-\int d^D x \,\LL(\phi^a,X^{bc})\,,
\ee
where $\phi^a=\phi^a(x^i)$ and $X^{bc}=-\phi^b_{,i} \phi^{c,i}/2$. Note that
\be
{T^i}_i=-2 \LL_{,X^{ab}}  X^{ab} + D \LL\,.
\ee
 
\subsection{Derrick's argument}

Let us apply Derrick's argument to the case of generic Lagrangians given by Eq. (\ref{gen}) (note that it holds only for finite energy configurations). Consider a function 
$E_\lambda$ defined by
\be\label{funcionalEnergia}
E_\lambda=-\int d^D x \,\LL(\phi^a_\lambda,X^{bc}_\lambda)\,, 
\ee
where $\phi^a_\lambda=\phi^a(\lambda x^i)$, $X^{bc}_\lambda=\phi^b_{\lambda,\mu} \partial^\mu \phi^{c,\mu}_\lambda/2$ and $\lambda$ is a real parameter. This function must satisfy the condition $E_1=E$. If we change the integration variable to $y^i=\lambda x^i$ we may rewrite Eq. (\ref{funcionalEnergia}) as
\be\label{funcionalEnergia2}
E_\lambda=-\int d^D y \,\lambda^{-D} \,\LL(\phi^a,\lambda^2 X^{bc})\,.
\ee
A static solution,  $\phi^a_s=\phi^a_s(x^i)$, must satisfy 
\bq\label{Dcondition}
\left[\frac{dE_\lambda}{d\lambda}\right]_{\lambda=1}&=&\int d^D x \,\left(D\LL-2\LL_{,X^{ab}} X^{ab}\right)\nonumber\\ 
&=& \int d^D x {T^i}_i=0\,,
\eq
and a necessary condition for the solution to be stable is that $E_\lambda$ has a minimum at $\lambda=1$. Hence, we 
require that
\bq
\label{D2condition}
\left[\frac{d^2E_\lambda}{d\lambda^2}\right]_{\lambda=1}&=&
\int d^D x \Big( -D({D+1})\LL \Big. \nonumber\\
&+&(4D-2) \LL_{,X_{ab}}X_{ab} \nonumber\\
&-&\Big. 4 \LL_{,X^{ab}X^{cd}}X^{ab} X^{cd}\Big) > 0\,.
\eq

\subsection{Specific example}

Consider a Lagrangian given by
\be
\label{nstanl}
\LL =X|X|^{n-1} - V(H)\,,
\ee
where $X={X^a}_a=\delta_{ab}   X^{a b}$, $H^2(r)=\delta_{ab} \phi^a \phi^b$ and $V(H)=V_0(H^2-1)^2$ (the units of mass were chosen so that the minimum of the potential is defined by $H=1$). Although it was argued in \cite{Babichev:2006cy} that the equation of motion (\ref{eqmotion}) becomes non-dynamical at $X=0$ for $n>1$, this problem may be resolved by adding a term $\epsilon X$ to the Lagrangian, where $\epsilon > 0$ may be arbitrarily small. Although this term formally leads to an infinite energy for any non-zero $\epsilon$, in physically realistic situations there is a cutoff at some length scale (for example, at the horizon). This means that the physical impact of the formal divergence is negligible if $\epsilon$ is small enough, fully justifying considering the above Lagrangian. Eq. (\ref{nstanl}) then implies that $E=K+U$ with 
\be\label{EGandEV1}
K=-\int d^Dx \, X|X|^{n-1} \, \,\,\,\, U=\int d^Dx \, V(\phi^b)\,.
\ee
The gradient energy of a maximally symmetric static solution in a $D+1$ Minkowski space-time is given by 
\be
K(r)=\frac{S_{D-1}}{2^n}\int_0^r \left((H_{,{\tilde r}})^2 + \frac{D-1}{{\tilde r}^2} H^2\right)^n {\tilde r}^{D-1} d {\tilde r} \,,
\label{erho2}
\ee
where $S_{D-1}=D\pi^{D/2}/\Gamma(D/2+1)$ and $\Gamma$ is the gamma function. At large distances from the defect core $H \to 1$ and consequently $E$ diverges when $r \to \infty$ for $n \le D/2$. For $n > D/2$ the total energy is finite. Consequently, it is possible to find global defects whose total energy does not diverge at spatial infinity by considering non-standard kinetic terms which localize most of the energy inside the defect core. However, if $D>1$ the energy density cannot have compact support due to the contribution of the gradient energy at arbitrary distances from the core (this is no longer true in the case of a local symmetry, as shown in \cite{Adam:2008rf,Bazeia:2010vb}).

Eq. (\ref{Dcondition}) implies that
\be
\label{Stderrick1}
(D-2n) K + D U=0\,.
\ee
This means that stable defects are possible in $D$ spatial dimensions if $n>D/2$. In the case of the standard Lagrangian given by Eq. (\ref{standardL}), Eq. (\ref{Dcondition}) leads to
\be
\label{Stderrick}
(D-2) K + D U=0\,,
\ee
where $K$ and $U$ are defined in Eq. (\ref{EGandEV}). If $V \ge 0$ for all $\phi^a$ then $U>0$. Consequently Eq. (\ref{Stderrick}) cannot be satisfied for $D \ge 2$, according to Derrick's argument. On the other hand, if $D=1$ then
\ben\label{D2condition1}
\left[\frac{d^2 E_\lambda}{d\lambda^2}\right]_{\lambda=1}=2 \int d x  V(\phi^a) = E >0\,,
\een
which signals the existence of stable static solutions with finite energy for $D=1$ (domain walls). A similar result also applies in the case of maximally symmetric static solutions of $D$-dimensional k-defects satisfying the ansatz (\ref{ans}), even if $D>1$, as long as $n>D/2$ (so that the total energy remains finite). Of course, this is no longer true for maximally symmetric D-dimensional domain walls with $D >1$ described by a real scalar field with $\phi(r)=H(r)$, since in that case there would be no topological constraint preventing $H(r=0)$ to move continuously from $-1$ to $1$, or vice-versa. Therefore, in the absence of external interactions, maximally symmetric D-dimensional domain walls are always unstable, regardless of the particular form of the scalar field Lagrangian.

The stability of standard global monopoles with respect to non-radial perturbations has been studied in \cite{1992NuPhB.375..665P,1989PhRvL..63.2158G,1991PhRvL..67.1173R,2000PhRvL..85.3091A,2002PhRvD..66h5019W}. Global monopoles were shown to be stable to infinitesimal normalizable perturbations with axial symmetry. However, it was further demonstrated that different topological sectors are separated by a finite energy barrier, independently of the details of the scalar field potential. In $3+1$ dimensions this feature is specific of standard global monopoles, whose energy is dominated by gradients far from core and does apply to the localized defects studied in the present paper.  In the case of stable localized defects different topological sectors are usually separated by infinite energy barriers.

Skyrmions with baryon number $B=1$ and Q-balls provide further examples of localized spherical symmetric defects in $3+1$ dimensions. Skyrmions \cite{1962NucPh..31..556S} are topological solutions of a lagrangian embodying chiral symmetry and have been widely used as a model for baryons while Q-balls \cite{1985NuPhB.262..263C} are stationary (not static) non-topological solitons whose stability is guaranteed by a conserved charge.

\subsubsection{\rm Asymptotic solution ($r \to 0$)}

In the case of maximally symmetric static solutions in a $D+1$ Minkowski space-time Eq.  (\ref{eqmotion}) becomes
\be
\label{eqmotion1}
(r^{D-1}\mathcal{L}_{,X}H_{,r})_{,r}=r^{D-1}\mathcal{L}_{,H}\,.
\ee
One may expand $\phi(r)$ near the origin as a polynomial in $r$ and take the lowest non-zero order $\phi(r) \propto r^m$, which is the dominant contribution in the $r \to 0$ limit. Using Eq.  (\ref{eqmotion}) it is simple to show that $m=1$ independently of the values of $n$ and $D$. Alternatively one could use the above ansatz $H({\tilde r})=A ({\tilde r}/r)^m$ (with $A>0$ and $m>0$, so that $K(r)$ is finite) in Eq.  (\ref{erho2}) to obtain
\be
K(r)=S_{D-1}\left(\frac{A^2}{2}\right)^n\frac{(m^2+D-1)^n}{2mn-2n+D} r^{-2n+D}\,.
\ee
$K(r)$ has a minimum at $m=1$, for any $n$ or $D$. Hence in the $r \to 0$ limit one has $\phi \propto r$, just as in the case of standard defects. 

\subsubsection{\rm Asymptotic solution ($r \to \infty$)}

Using Eq. (\ref{eqmotion1}) it is also possible to obtain the behavior of the solution at large distances from the defect core. If $D>1$ then $H \sim 1 - Br^{-2n}+O(r^{-2(n+1)})$ with $B>0$ so that the gradient and potential energy densities are proportional to $r^{-2n}+O(r^{-2(n+1)})$ and $r^{-4n}+O(r^{-4(n+1)})$, respectively. Hence, for $D>1$ and $n>1$ the gradient energy density dominates over the potential energy density far away from the core and consequently the total density  is also proportional to $r^{-2n}+O(r^{-2(n+1)})$. This confirms the result that, for $n >1$, the total energy, $E$, is finite if $n>1$. However, for $D>1$ global defects can only become compact-like in the limit of very large $n$. 

\subsection{Cosmological implications}

In the case of maximally symmetric solutions Eq. (\ref{Dcondition}) implies that all the spatial components of the energy-momentum tensor vanish. On the other hand, if we consider $N>D$, with the scalar field multiplet being independent of the additional $N-D$ space-time coordinates, then $T_{ii}=-T_{00}$. This means that a network of static localized defects will have an (average) equation of state given by
\be\label{funcionalEqSt}
{\mathcal P}=-\frac{N-D}{N}\rho\,,
\ee
independently of the specific Lagrangian of the model (here $\rho$ and $\mathcal P$ represent the average energy density and pressure associated with the defect network). However, if the defects have a non-zero root mean square
velocity, $v$, then the (average) equation of state parameter becomes
\be\label{funcionalEqSt1}
w=\frac{\mathcal P}{\rho}=-\frac{N-D}{N}+\frac{N-D+1}{N}v^2\,,
\ee
so that $w \to 1/N$ when $v \to 1$ \cite{Avelino:2008mv}. 

If we ignore radiation effects then the dynamics of thin p-branes (with $p=N-D$) is independent of their internal structure along the spherically symmetric subspace of dimension $D$, assuming that the brane is featureless along the $p$ space-like coordinates parallel to it (this was explicitly demonstrated in \cite{Avelino:2008ve,2010PhRvD..81h7305S} for domain walls with $D=1$ but it is also true for $D>1$). 

The case with $N=3$, $D=2$ and $n>1$ is that of localized global strings whose dynamics, if their thickness is much smaller than their curvature radius, is governed by the Nambu-Goto action \cite{1971PThPh..46.1560G} rather than the Kalb-Ramond one \cite{1974PhRvD...9.2273K}, just as in the case of local strings. Hence, although some of the decay channels are different for local and localized global strings, the corresponding cosmological implications are expected to be quite similar. 

The case with $N=3$, $D=3$ and $n>1$ is that of localized global monopoles. If $n=1$, in a scaling regime, there are typically only a few monopoles per Hubble volume (see, for example, \cite{Bennett:1990xy,Pen:1993nx}). However, localized global k-monopoles, if they are sufficiently apart from one another, interact very little and are slow down by the expansion of the Universe, their velocity $v$ being proportional to $a^{-1}$ where $a$ is the cosmological scale factor, thus providing an energy source with $p \sim 0$. Localized k-monopoles  are therefore an interesting dark matter candidate (their energy density cannot exceed $\sim 20 \%$ of the background density today). At the present time, the constraints on the energy density of standard global monopoles are much tighter \cite{1997PhRvL..79.1611P}. However, one should bear in mind that the average energy density of a standard global monopole network scales linearly with background density in the radiation and matter eras while localized monopoles tend to dominate the energy density of the universe during the radiation era (the long range interactions are much weaker in the later than in the former case).  Consequently, at early times the constraints on the energy density of localized monopoles turn out to be much more stringent than for standard global monopoles. Also note that, at the present time, constraints on local magnetic monopoles are much stronger (e.g. Parker Bound \cite{1970ApJ...160..383P}) compared to the case of localized global k-monopoles which do not source a magnetic field.

\subsection{Graphene}

Graphene is a two-dimensional honeycomb lattice of carbon atoms which can be viewed as the superposition of two identical triangular sublattices, each possessing two Dirac points. In $2+1$ space-time dimensions, the Dirac Hamiltonian for massless fermions in the presence of Yukawa coupling with a complex scalar field $\phi=\phi(x,y)$ can be written as
\begin{equation}
H=\left(\begin{array}{c c c c}
0&i\partial_{z}&\phi&0\\
i\partial_{z}&0&0&\phi\\
\bar\phi&0&0&i\partial_{\bar z}\\
o&\bar\phi&i\partial_{\bar z}&0
\end{array}\right)
\end{equation}
where $z=x+iy$, $x$ and $y$ are two-dimensional Cartesian coordinates and the bar denotes complex conjugation. In this case the Dirac matrices are $4\times4$ because there are four degrees of freedom, two for each one of the two sublattices of graphene. The phase of the complex scalar field may vary in space and be associated to vortices leading to charge fractionalization \cite{chamon}. However, the energy of standard global vortices is divergent. For this reason, in \cite{jackiw} the authors proposed a way to circumvent the situation, making the vortices local, through the presence of a gauge field. Although the suggestion is interesting, it is not yet fully understood what could be responsible for such dynamical gauge field. In the present work we have modified the global scalar field dynamics in a way which allows for finite energy vortex configurations by considering a Lagrangian with a non-standard kinetic term. Thus, our model may provide an alternative to the idea proposed in \cite{jackiw}, without the need for the gauge field.

\section{Conclusions}

In summary, in this paper we have demonstrated the existence of localized static global k-defects of arbitrary dimensionality by considering maximally symmetric solutions described by a simple extension of a standard Lagrangian in a $D+1$ dimensional Minkowski space-time. We have shown that the density profiles can change dramatically with respect to the standard case, specially at large distances from the defect centre, leading to static solutions whose energy does not diverge at spatial infinity. By applying Derrick's argument to maximally symmetric defects described by generic scalar field Lagrangians we obtained a model independent relation between the various components of the defect energy-tensor and determined the (averaged) equation of state of a network of localized global defects. 
We have shown that the defects are stable against radial stretching and argued in favor of the stability against angular deformations. A detailed analysis of the linear stability against angular deformations will be left for future work. Our findings have profound implications for the evolution of k-defect networks. A particularly interesting case is that of localized global monopoles which could be a relevant dark matter source.

Our model may also be used to obtain the fractionalization of charge in various physical systems, the graphene being perhaps the example of major current interest. In \cite{chamon} a mechanism leading to the fractionalization of  the electronic charge was presented. There, however, the author considers a global scalar field with vortex excitations of divergent energy. In this paper we have shown that a finite vortex energy may be obtained by considering a non-standard dynamics for the scalar field, circumventing the need for the gauge field proposed in \cite{jackiw}.

\begin{acknowledgments}

The authors would like to thank CAPES, CNPq (Brasil) and FCT (Portugal) for partial support.

\end{acknowledgments}


\bibliography{globalRev}

\end{document}